\documentclass{article}
\usepackage[utf8]{inputenc}
\usepackage{amsmath}
\usepackage{nth}
\usepackage{titlesec}
\usepackage{graphicx}
\usepackage{multicol}
\usepackage{floatrow}
\usepackage{caption}
\usepackage{mathtools}
\usepackage{subfig}
\PassOptionsToPackage{hyphens}{url}\usepackage{hyperref}
\usepackage{adjustbox}
\title{\textbf{A Pan-Cancer and Polygenic Bayesian Hierarchical Model for the Effect of Somatic Mutations on Survival}}

\date{}

\author{Sarah Samorodnitsky$^{1}$, Katherine A. Hoadley$^{2}$, and Eric F. Lock$^{1}$ \\ \\
\small{$^1$Division of Biostatistics, School of Public Health, University of Minnesota} \\ \\
\small{$^2$Department of Genetics,  Computational Medicine Program, University of North Carolina }\\}

\begin{document}


\maketitle

\begin{abstract}
    We built a novel Bayesian hierarchical survival model based on the somatic mutation profile of patients across 50 genes and 27 cancer types. The pan-cancer quality allows for the model to “borrow” information across cancer types, motivated by the assumption that similar mutation profiles may have similar (but not necessarily identical) effects on survival across different tissues-of-origin or tumor types. The effect of a mutation at each gene was allowed to vary by cancer type while the mean effect of each gene was shared across cancers. Within this framework we considered four parametric survival models (normal, log-normal, exponential, and Weibull), and we compared their performance via a cross-validation approach in which we fit each model on training data and estimate the log-posterior predictive likelihood on test data.  The log-normal model gave the best fit, and we investigated the partial effect of each gene on survival via a forward selection procedure. Through this we determined that mutations at \emph{TP53} and \emph{FAT4} were together the most useful for predicting patient survival. We validated the model via simulation to ensure that our algorithm for posterior computation gave nominal coverage rates. The code used for this analysis can be found at \url{http://github.com/sarahsamorodnitsky/Pan-Cancer-Survival-Modeling}, and the results are at \url{http://ericfrazerlock.com/surv_figs/SurvivalDisplay.html}.
    
    \vspace{12pt}
\end{abstract}

\section{Introduction}
\label{intro}
The recently completed Cancer Genome Atlas (TCGA) program has provided a comprehensive molecular characterization of 33 cancer types from over 10,000 patients, and the database remains a valuable public resource \cite{hutter2018cancer}.  The different cancer types are generally defined by their tissue-of-origin. The project has revealed striking heterogeneity in the genomic and molecular profiles across patients within each cancer type.  This heterogeneity is presented in several flagship publications (e.g., see \cite{cancer2012comprehensive,cancer2014comprehensive,verhaak2010integrated}) in the form of molecularly distinct subtypes, and these subtypes often correlate with clinical endpoints.  However, leveraging molecular heterogeneity for personalized risk prediction on a more granular scale is limited by the number of patients with reliable clinical data for each cancer type \cite{liu2018integrated}.  Moreover, the effects of individual molecular biomarkers on survival or other clinical outcomes are often small, and thus predictive analyses within a single type of cancer are underpowered \cite{yuan2014assessing}. 

In 2013, TCGA began the Pan-Cancer Analysis Project, motivated by the observation that ``cancers of disparate organs reveal many shared features, and, conversely, cancers from the same organ are often quite distinct" \cite{weinstein2013cancer, hoadley2018cell}.  This initiative has resulted in several studies across multiple cancer types that have
revealed important shared molecular alterations for somatic mutations \cite{kandoth2013mutational}, copy number \cite{zack2013pan}, mRNA \cite{hoadley2014multiplatform}, and protein abundance \cite{akbani2014pan}.  If these potential biomarkers have similar clinical effects across multiple types of cancer, then predictive models that are estimated using pan-cancer data will have more power than models that are fit separately for each type of cancer.  The TCGA Pan-Cancer clinical data resource (TCGA-CDR) \cite{liu2018integrated} has facilitated pan-cancer models by curating and standardizing available data for four clinical outcomes (overall survival, disease-specific survival, disease-free interval or progression-free interval) across all 33 TCGA cohorts.  

	 We propose and implement a novel Bayesian hierarchical framework to predict survival from multiple molecular predictors that are shared across multiple cancer types.  An important feature of this approach is that it allows for borrowing information across models for each cancer type under the assumption that shared molecular predictors are likely to have a similar effect on survival prognosis, but it also allows sufficient flexibility for the same biomarker to have different effects depending on the type of cancer.  Moreover, the Bayesian framework provides a principled way to incorporate prior information from other studies or cohorts, which is well-motivated given the vast body of literature on molecular biomarkers in cancer.  

	Using our proposed framework, we developed a pan-cancer predictive model for overall survival (OS), using the somatic mutation profile of each tumor as the primary predictors of interest.   We used OS because it is unambiguously defined and is available for almost all types of cancer, despite short follow-up times \cite{liu2018integrated}.  We focus on somatic mutations because they play a critical role in the development of many cancer types \cite{martincorena2015somatic}, are available for all cohorts in the TCGA database, and are straightforward to compare across different tissues-of-origin.  A pan-cancer analysis of somatic mutations across 12 different cancer types from TCGA revealed several genes that have frequent mutations across multiple types of cancer \cite{kandoth2013mutational}.  Their analysis also considered the marginal effect of each gene on OS, via cox proportional hazards models for (1) each cancer type separately and (2) for a fully joint model with all cancer types together; our approach aims to compromise between these two strategies.  For our application we consider the somatic mutation status of 50 genes, and survival data for 5698 patients comprising 27 different types of cancer.  We evaluate and compare several different potential models via a robust cross-validation approach to assess predictive accuracy for this dataset.
	
	The rest of this article is organized as follows.  In the Methods section, we discuss how the data were collected and our approach to filtering out cancer types and genes. We also describe our modeling framework, how we selected the survival distribution we chose to use in our analysis, and our Gibbs sampling algorithm to calculate the posteriors of the model parameters. In the Results section, we discuss the results of our model selection procedure and our approach to determining which genes were most predictive of survival. Using the results from these two processes, we show the resulting credible interval estimates of our parameters and display survival curves. In the Discussion section, we discuss our results and future work that builds on this study for pan-cancer survival modeling.

\section{Methods}
\label{methods}
\subsection{Data Acquisition and Processing} 
We acquired clinical data for each patient via the Cancer Genome Atlas Clinical Data Resource (TCGA-CDR)\cite{liu2018integrated}, which includes data for 33 cancer types and over 11,160 patients. We acquired genome wide somatic mutation data through the TCGA2STAT package for R\cite{wan2015tcga2stat}, which gathers data from the Broad Institute GDAC Firehose.  
The curated mutation data were available for $27305$ genes and $5793$ patients with a binary indicator for whether there was a somatic non-silent mutation ($0$=no, $1$=yes) within the coding region of each gene. 

We first 
matched the observations in the CDR dataset with the mutation data obtained through TCGA2STAT. If any patients were present in one dataset but not the other, that observation was removed from the study. We also removed any observations that had a negative survival time, a survival time entered as $0$, or who were missing both a survival time and an entry for time-to-last-contact. We chose to eliminate five cancer types from our study due to high ($>90\%$) censoring rates, meaning patients survived longer than the duration of the study and their outcome status is unknown. 
These five types were Pheochromocytoma and Paraganglioma (PCPG), Prostate adenocarcinoma (PRAD), Testicular Germ Cell Tumors (TGCT), Thyroid carcinoma (THCA), and Thymoma (THYM). 
The TCGA-CDR also caution against using OS as an endpoint for these cancer types, due to the lack observed survival events.  One other cancer type, Mesothelioma (MESO), did not have any somatic mutation data available through the TCGA2STAT pipeline, so it was also omitted from our analysis. 

The following 27 cancer types remained in our study: Adrenocortical carcinoma (ACC), Bladder Urothelial Carcinoma (BLCA), Breast invasive carcinoma (BRCA), Cervical squamous cell carcinoma and endocervical adenocarcinoma (CESC), Cholangiocarcinoma (CHOL), Colon adenocarcinoma (COAD), Lymphoid Neoplasm Diffuse Large B-cell Lymphoma (DLBC), Esophageal carcinoma (ESCA), Glioblastoma multiforme (GBM), Head and Neck squamous cell carcinoma (HNSC), Kidney Chromophobe (KICH), Kidney renal clear cell carcinoma (KIRC), Kidney renal papillary cell carcinoma (KIRP), Acute Myeloid Leukemia (LAML), Brain Lower Grade Glioma (LGG), Liver hepatocellular carcinoma (LIHC), Lung adenocarcinoma (LUAD), Lung squamous cell carcinoma (LUSC), Ovarian serous cystadenocarcinoma (OV), Pancreatic adenocarcinoma (PAAD), Rectum adenocarcinoma (READ), Sarcoma (SARC), Skin Cutaneous Melanoma (SKCM), Stomach adenocarcinoma (STAD), Uterine Corpus Endometrial Carcinoma (UCEC), Uterine Carcinosarcoma (UCS), Uveal Melanoma (UVM). 

We filtered the genes based on average mutation rate across the 27 cancer types, selecting the top 50 mutated genes. In this way, each cancer type was weighted the same in calculating the mean mutation rate. 
This also ensured that each gene would be represented across 
most cancers, not just within a few. As a result, certain genes which are highly mutated in particular cancers but not in others were excluded. The genes we incorporated in our study were \emph{ABCA13, AHNAK2, APC, APOB, ARID1A, CSMD1, CSMD2, CSMD3, DMD, DNAH11, DNAH5, DNAH7, DNAH8, DNAH9, DST, FAT1, FAT3, FAT4, FLG, FRG1B, GPR98, HMCN1, KRAS, LRP1B, LRP2, MACF1, MLL2, MLL3, MUC16, MUC17, MUC2, MUC4, MUC5B, NEB, OBSCN, PCDH15, PCLO, PIK3CA, PKHD1L1, PTEN, RYR1, RYR2, RYR3, SPTA1, SYNE1, TP53, TTN, USH2A, XIRP2,} and \emph{ZFHX4}. The average rate of mutation for these genes across the 27 cancer types is summarized in Table~\ref{tab:threeparts}. In total, we used mutation data from 5698 patients. 

\begin{figure}[H]
  \includegraphics[width=0.75\linewidth]{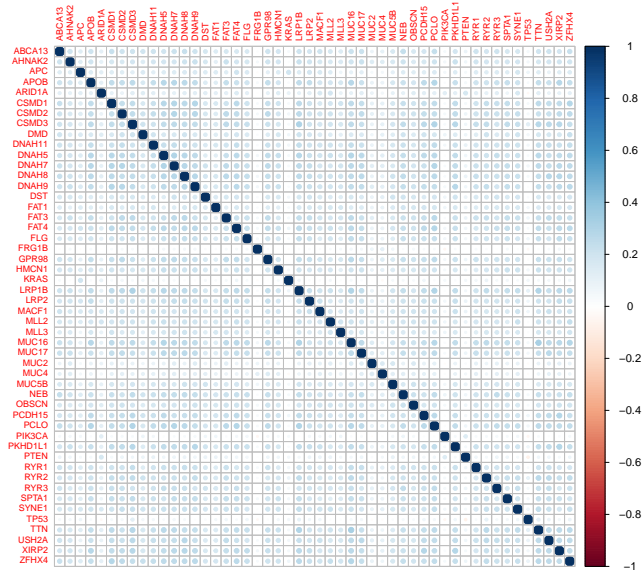}
  \caption{Correlation plot for correlations between somatic mutation statuses across tissue types}
  \label{fig:Correlation_Plot}
\end{figure}

We considered the correlation of mutation status between genes across all cancers, for exploratory purposes and to investigate potential issues of multicollinearity for polygenic models.  
Figure~\ref{fig:Correlation_Plot} shows a pairwise correlation plot for all genes considered using mutation status across all patients included in this study. The Pearson correlation coefficients between 
genes were uniformly positive but relatively week, ranging from $r=-0.07$ to $r=0.32$.  The positive correlations were expected as the total mutation burden can vary across patients; however, the relative weakness of the correlations suggests that each individual gene may provide unique information and multicollinarity is not a concern.  These correlations may change considerably if one were to consider only a single type of cancer or a subset of related cancers.    

\begin{table}[H]
   \caption{Summary of Genes and Average Mutation Rate}\label{tab:threeparts}
   \centering
   \begin{tabular}{cc}
     Gene & Mutation Rate \\ \hline
     \emph{TP53} & 0.357  \\
     \emph{TTN} & 0.296  \\
     \emph{MUC16} & 0.192  \\
     \emph{MUC4} & 0.124  \\
     \emph{PIK3CA} & 0.116  \\
     \emph{CSMD3} & 0.116  \\
     \emph{LRP1B} & 0.115  \\
     \emph{SYNE1} & 0.110  \\
     \emph{KRAS} & 0.108  \\
     \emph{FLG} & 0.105  \\
     \emph{RYR2} & 0.104 \\
     \emph{USH2A} & 0.098 \\
     \emph{PCLO} & 0.095 \\
     \emph{APC} & 0.094 \\
     \emph{DNAH5} & 0.093 \\
     \emph{MUC5B} & 0.089 \\
     \emph{FAT4} & 0.088 \\
     \emph{OBSCN} & 0.088 \\
     \emph{CSMD1} & 0.084 \\
     \emph{MUC17} & 0.084 \\
     \emph{ZFHX4} & 0.084 \\
     \emph{HMCN1} & 0.083 \\
     \emph{ARID1A} & 0.080 \\
     \emph{GPR98} & 0.080 \\
     \emph{FAT3} & 0.077 \\
   \end{tabular}
   \hspace{1em}
   \begin{tabular}{cc}
     Gene & Mutation Rate \\ \hline
     \emph{LRP2} & 0.077  \\
     \emph{XIRP2} & 0.077  \\
     \emph{AHNAK2} & 0.077  \\
     \emph{MLL2} & 0.077  \\
     \emph{SPTA1} & 0.076  \\
     \emph{APOB} & 0.076  \\
     \emph{PTEN} & 0.076  \\
     \emph{MLL3} & 0.075  \\
     \emph{FRG1B} & 0.074  \\
     \emph{PKHD1L1} & 0.073  \\
     \emph{DST} & 0.070  \\
     \emph{DMD} & 0.070  \\
     \emph{RYR3} & 0.070  \\
     \emph{MUC2} & 0.070  \\
     \emph{NEB} & 0.067  \\
     \emph{RYR1} & 0.066  \\
     \emph{MACF1} & 0.066  \\
     \emph{PCDH15} & 0.066  \\
     \emph{ABCA13} & 0.066  \\
     \emph{DNAH9} & 0.064  \\
     \emph{FAT1} & 0.064  \\
     \emph{DNAH8} & 0.063 \\
     \emph{DNAH11} & 0.062  \\
     \emph{CSMD2} & 0.061  \\
     \emph{DNAH7} & 0.060  \\
   \end{tabular}
\end{table}

\subsection{Model}
We propose a Bayesian hierarchical model for patient survival that incorporates binary mutation status variables and age across 27 cancer types. The multi-layer nature of our model allows the effect of a mutation at each gene to vary by cancer type, 
while simultaneously inferring the mean and variance of these effects.  
Thus, the model facilitates the borrowing of information across cancer types by shrinking the estimated effects towards a common mean.  
Our model can also accommodate censored observations, 
as discussed in the following subsection. We use the following notation in our framework: $y_{ij}$ is the (potentially censored) survival time for patient $j$ in cancer type $i$, $j = 1, \dots, n_i$, $i = 1, \dots, 27$.  
$x_{ijp}$ is the centered age if $p=1$ and is the mutation status for gene $p-1$, $p=2, \dots, 51$. 
We consider four different likelihood models for survival:  
\begin{align*}
    y_{ij} &\sim \mbox{Normal}(\lambda_{ij}, \sigma^2) \\
    y_{ij} &\sim \mbox{Log-normal}(\lambda_{ij},\sigma^2) \\
    y_{ij} &\sim \mbox{Exponential}(\frac{1}{\lambda_{ij}}) \\
    y_{ij} &\sim \text{Weibull}(\alpha, \frac{1}{\lambda_{ij}}).
\end{align*}
\\
For each likelihood we consider a hierarchical linear model for $\lambda_{ij}$:
\[\lambda_{ij} = \beta_{i0} + x_{ij1}\beta_{i1} + x_{ij2}\beta_{i1} + \dots + x_{ij51}\beta_{i51}\] where $\beta_{ip}$
is the linear effect of age if $p=1$ and mutation at gene $p-1$ on survival if $p = 2, \dots, 51$ for a patient with cancer type $i$. 
This approach extends commonly used parametric Bayesian survival models \cite{ibrahim2013bayesian}, and to complete the Bayesian framework we specify prior distributions for the unknown parameters.     For the Normal and Log-normal models, we used an Inverse-Gamma$(0.01, 0.01)$ prior distribution for the residual variation of survival times within each cancer type, $\sigma^2$. 
For the Weibull model, we used a Uniform$(0,5)$ prior for the shape parameter, $\alpha$. 
Under the assumption that a mutation at one gene or an increase in one year of age may affect survival differently depending on the 
type of cancer, the linear effect of age and each mutation, $\beta_{ip}$, was assumed to vary by cancer type. We assume \[\beta_{ip} \sim \mbox{Normal}(\tilde\beta_p, \lambda_p^2) \; \text{ for } p=0,\hdots,51.\] where $\tilde\beta_p$ is the mean effect on survival across cancer types and $\lambda_p$ describes the extent to which the effects vary across the different types.  Thus, $\tilde{\beta}_{i0}$ gives the mean intercept and $\beta_{i0}$ is an intercept that describes the baseline survival for type $i$.  Similarly, $\tilde\beta_p$ is the average effect on survival for age if $p=1$ and for mutation at a gene if $p=2, \dots, 51$. For all coefficients and all models, we gave the mean effects independent and diffuse normal priors: $\tilde\beta_p \sim \mbox{Normal}(0, 10000^2)$.  The parameters $\lambda_p^2$ are important because they indicate the degree of effect heterogeneity across cancer types; we used Inverse-Gamma $(0.01, 0.01)$ priors for each $\lambda_p^2$.  

\subsection{Parameter Estimation}
Depending on the survival model, we employed a different approach to infer the posterior for model parameters. The log-normal and normal models were fit in R using an in-house Gibbs sampler. The sampler is described below however more details can be found in Appendix 1. 

\begin{enumerate}
    \item Initialize $\tilde\beta_p^{(0)}$, $\lambda_p^{2(0)}$, $\sigma^{2(0)}$. Initialize all censored observations at their time-of-last-contact. \\ For samples $t = 1, ..., 20000$, repeat the following steps:
    \item Draw $\beta_{ip}^{(t)}$ from $P(\beta_{ip}^{(t)}|\tilde\beta^{(t-1)}, \lambda^{2(t-1)}, \sigma^{2(t-1)})$ for $i=1, \dots, 27$ and $p = 0, \dots, 51$.
    \item Draw $\lambda_p^{2(t)}$ from $P(\lambda_p^{2(t)}|\beta_{ip}^{(t)}, \tilde\beta^{(t-1)})$ for $p=0, \dots, 51$.
    \item Draw $\tilde\beta_p^{(t)}$ from $P(\tilde\beta_p^{(t)}|\beta_{ip}^{(t)}, \lambda_p^{2(t)})$ for $p=0, \dots, 51$.
    \item Draw $\sigma^{2(t)}$ from $P(\sigma^{2(t)}|\beta_{ip}^{(t)})$ for for $i=1, \dots, 27$ and $p = 0, \dots, 51$.
    \item Generate survival times for censored observations using $\beta_{i1}^{(t)}, ..., \beta_{ip}^{(t)}, \sigma^{2(t)}$
    \begin{itemize}
        \item If assuming the data follows a normal distribution, generate survival times for censored observations from a normal distribution with mean $\beta_{i0}^{(t)} + \beta_{i1}^{(t)}x_{ij1} + \dots + \beta_{i51}^{(t)}x_{ij51}$ and variance $\sigma^{2(t)}$ that is truncated at the time-of-last-contact for observation $ij$.  
        \item If assuming the data follows a log-normal distribution, generate survival times from a normal distribution with mean $\beta_{i0}^{(t)} + \beta_{i1}^{(t)}x_{ij1} + \dots + \beta_{i51}^{(t)}x_{ij51}$ and variance $\sigma^{2(t)}$ that is truncated at the log of the time-of-last-contact for observation $ij$. 
    \end{itemize}
\end{enumerate}

We ran the sampler for 20000 iterations and used a 10000 iteration burn-in to ensure convergence of the parameters. In the Appendix 2 of this article, we describe how we validated this model fitting procedure. 

Posterior samples for the exponential and Weibull models were obtained using the Just Another Gibbs Sampler (JAGS) software\cite{plummer2003jags}.  For these models, we ran the sampler for the same number of iterations and employed the same burn-in as for the normal and log-normal models above. In all calculations based on the posteriors of our Gibbs sampler and our JAGS models, we thinned by every 10th iteration to speed up computing time and memory efficiency.  

\subsection{Model Selection}
To assess which of the normal, log-normal, exponential, and Weibull models was the best fit for the data, we calculated the log out-of-sample posterior predictive likelihood in a 5-fold cross validation procedure, as described below. 

Consider the $k$th training-test partition of the data, $k = 1, \dots, 5$ such that $\vec{Y} = \{\vec{Y}_k^{\text{train}}, \vec{Y}_k^{\text{test}} \}$ . Let $p(y|X)$ be the probability distribution for survival time. On each training fold, we fit the model and generated posterior samples for each parameter. For each posterior sample after burn-in, we computed
\[
P(\vec{Y}^{\text{test}}| \Theta_o^t) = \prod_{\substack{(i,j) \\ \text{uncensored}}} p(y_{ij}|\Theta_o^t) \prod_{\substack{(i,j) \\ \text{censored}}} \Pr(y_{ij} > y_{ij}^c \mid \Theta_o^t)
\]
where $\Theta_o^t$ is a vector of all the $t$th iteration posterior samples for the parameters of the probability distribution of survival and $y_{ij}^c$ is the censor time for the $j$th patient in the $i$th cancer type. After computing this quantity for each iteration, we computed an estimate of the out-of-sample posterior predictive likelihood:

\[
\int P(\vec{Y}^{\text{test}}|\Theta_0) P(\Theta_0|\vec{Y}^{\text{train}})d\Theta_0 \approx 
\frac{1}{T} \sum_{t=1}^T P(\vec{Y}^{\text{test}}|\Theta^t)
\]
where T is the number of sampling iterations after burn-in and thinning. As stated previously, we chose to thin by every 10th iteration to ease computing time. As a result, this value was calculated based on 1000 Gibbs sampling iterations. After calculating the log-posterior predictive likelihood on each test fold, we took the average likelihood and compared the four models.

\subsection{Forward Selection}
To assess the partial improvement of each gene in predicting survival, we constructed a forward selection approach that would allow us to see which genes were most important in predicting survival across all cancer types. In this way, we were able to determine the relative importance of each gene, and achieve a more parsimonious predictive model. For our forward selection approach, we used the same out-of-sample log-posterior likelihood metric described in the previous section on model comparison. Our forward selection method proceeded as follows: 

\renewcommand{\theenumii}{\arabic{enumii}}
\begin{enumerate}
    \item Calculate the average log-posterior predictive likelihood for the null model (with age and cancer type intercepts, but no genes) fit on each of the 5-folds.
    \item For each gene, consider the model with only the intercept, age, and that gene included. Calculate the log-posterior predictive likelihood under this model using 5-fold cross validation. 
    \begin{enumerate}
        \item Select the gene that produced the model with the highest log-posterior predictive likelihood. Call this model $M_1$.
        \item Compare the likelihood for this model with the null model; if the likelihood has increased, proceed by adding this gene to the model. Otherwise, stop. 
    \end{enumerate}
    \item For each remaining gene, add that gene separately to model $M_1$ and calculate the resulting mean log-posterior predictive likelihood using 5-fold cross validation. 
    \begin{enumerate}
        \item Select from the resulting two gene models the gene that maximized the log-posterior likelihood. Call this model $M_2$
        \item Compare the log-posterior likelihood for $M_2$ to $M_1$. If the likelihood has increased, add the new gene to the model and proceed. Otherwise, stop. 
    \end{enumerate}
    \item Continue until the log-posterior predictive likelihood ceases to increase. At this point, the final model has been found. 
\end{enumerate}

Once a final model had been found, we investigated the effects of each mutation on survival through credible interval plots and survival curves, described further in the Results section. 

\section{Results} 
\subsection{Model Selection} 
 The results of the comparison described in the Methods section are shown in Table~\ref{tab:cv_comparison}. We found that the log-normal model had the highest out-of-sample log-posterior likelihood out of the Normal, Exponential, and Weibull models. We assumed a log-normal distribution of survival for the remainder of our investigation. \\

\begin{table}[H]
\centering
\caption{Out-Of-Sample Posterior Predictive Likelihood}
\label{tab:cv_comparison}
\begin{tabular}{cc}
Model       & Log Posterior Likelihood \\ \hline
Log-Normal  & -3667.139                \\ \hline
Normal      & -3918.152               \\ \hline
Exponential & -7123.648                \\ \hline
Weibull     & -7212.59                \\ \hline
\end{tabular}
\end{table}

In this model, the coefficients are inferred by borrowing information across cancer types. However, we considered an analogous log-normal model in which the coefficients were inferred independently for each cancer type to compare with the model we selected here. We used the same uninformative prior on each coefficient as the one assumed for $\tilde{\beta}: \beta_{ij} \sim N(0,10000^2)$. However, our in-house Gibbs sampler under this model failed to converge after 30000 iterations for several coefficients. This demonstrates a drawback to not borrowing across cancer types as our proposed model does. 

\subsection{Forward Selection}
Table~\ref{tab:forward_selection} displays the mean log-posterior predictive likelihoods for each step in the forward selection procedure. Every model is adjusted for patient age. 
\begin{table}[H]
\centering
\caption{Results from Forward Selection Procedure}
\label{tab:forward_selection}
\begin{tabular}{cc}
Covariates in Model & Mean Log Posterior Likelihood \\ \hline
Age, No Genes  & -5014.895               \\ \hline
Age, \emph{TP53}      & -1009.63                  \\ \hline
Age, \emph{TP53}, \emph{FAT4} & -1009.135               \\ \hline
Age, \emph{TP53}, \emph{FAT4}, \emph{DNAH5}   & -1009.179     \\ \hline
\end{tabular}
\end{table} 

The log-posterior likelihood for the null model, meaning the model with only age as a predictor, was $-5014.895$. The model with \emph{TP53} added yielded a dramatic improvement, with  a log-posterior likelihood value of $-1009.63$. The next gene to be added was \emph{FAT4}, yielding a log posterior likelihood of $-1009.135$. At this point, the posterior likelihood stopped improving. \emph{DNAH5} was last to be added, and the model with \emph{TP53}, \emph{FAT4}, and \emph{DNAH5} as predictors led to a log-posterior likelihood of $-1009.179$. We validated convergence of the forward selection procedure by running the process several times to ensure we obtained the same result. Across all 27 cancers, \emph{TP53} was mutated in an average of 36.9\% of patients, the highest of all genes, and \emph{FAT4} in 8.8\% (Table~\ref{tab:threeparts}). The appearance of \emph{TP53} here is not surprising (see the Discussion section). However, we note that after the inclusion of \emph{TP53}, the improvement in likelihood by \emph{FAT4} was marginal. This final model from our forward selection procedure served as our basis of exploration for subsequent analysis.

We also used our forward selection approach without including \emph{TP53} as a potential covariate to see what genes would be added. Our results were as follows: 

\begin{table}[H]
\centering
\caption{Results from Forward Selection Procedure Without \emph{TP53}}
\label{tab:forward_selection}
\begin{tabular}{cc}
Covariates in Model & Mean Log Posterior Likelihood \\ \hline
Age, No Genes  & -5014.895                 \\ \hline
Age, \emph{APOB}      & -1011.321                 \\ \hline
Age, \emph{APOB}, \emph{ARID1A} & -1011.694               \\ \hline
\end{tabular}
\end{table} 

Without including \emph{TP53} as a possible covariate, the first gene to be added was \emph{APOB}, leading to a log-posterior likelihood of $-1011.321$. The addition of \emph{ARID1A} on top of \emph{APOB} led to the highest log-posterior likelihood of all genes at $-1011.694$ though this metric ceased to increase at this point. Of the patients in our study, 7.6\% had a mutation at \emph{APOB}. 

\subsection{Credible Intervals for Model Coefficients}
To understand the magnitude and direction of the partial effect of age and a mutation at a gene on patient survival, we computed and visualized the $95\%$ credible interval based on posterior samples for each $\beta_{ip}$. 
The intervals we show here were calculated from the multivariate log-normal model resulting from the forward selection procedure, with cancer type intercepts, age, \emph{TP53}, and \emph{FAT4} included as predictors. The credible intervals for each $\beta_{ip}$ can be found in Table~\ref{tab:cred_intervals}. 

\begin{figure}[H]
\centering
\sidesubfloat[]{\includegraphics[width=0.45\textwidth]{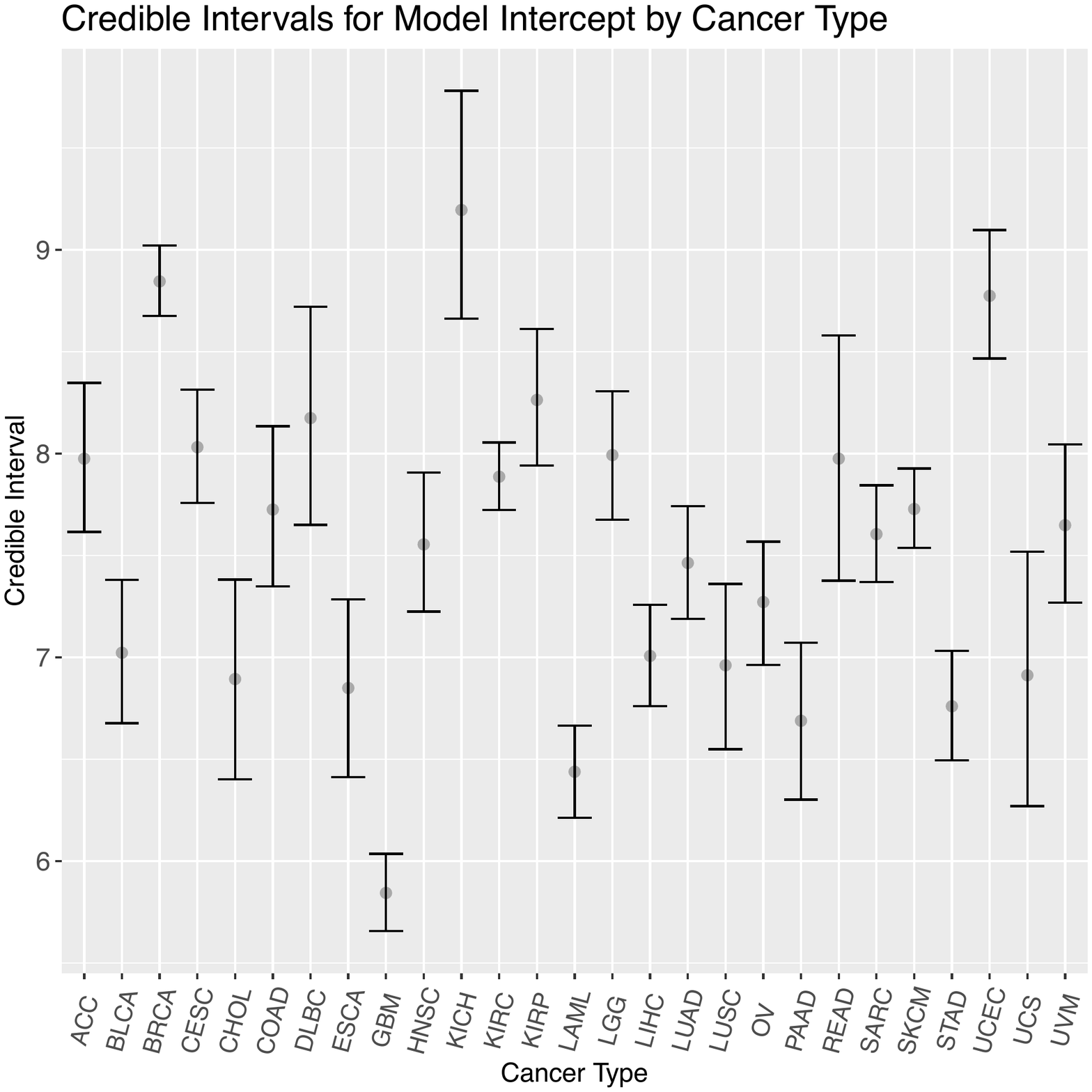}  \label{fig:a}}
\hfil
\sidesubfloat[]{\includegraphics[width=0.45\textwidth]{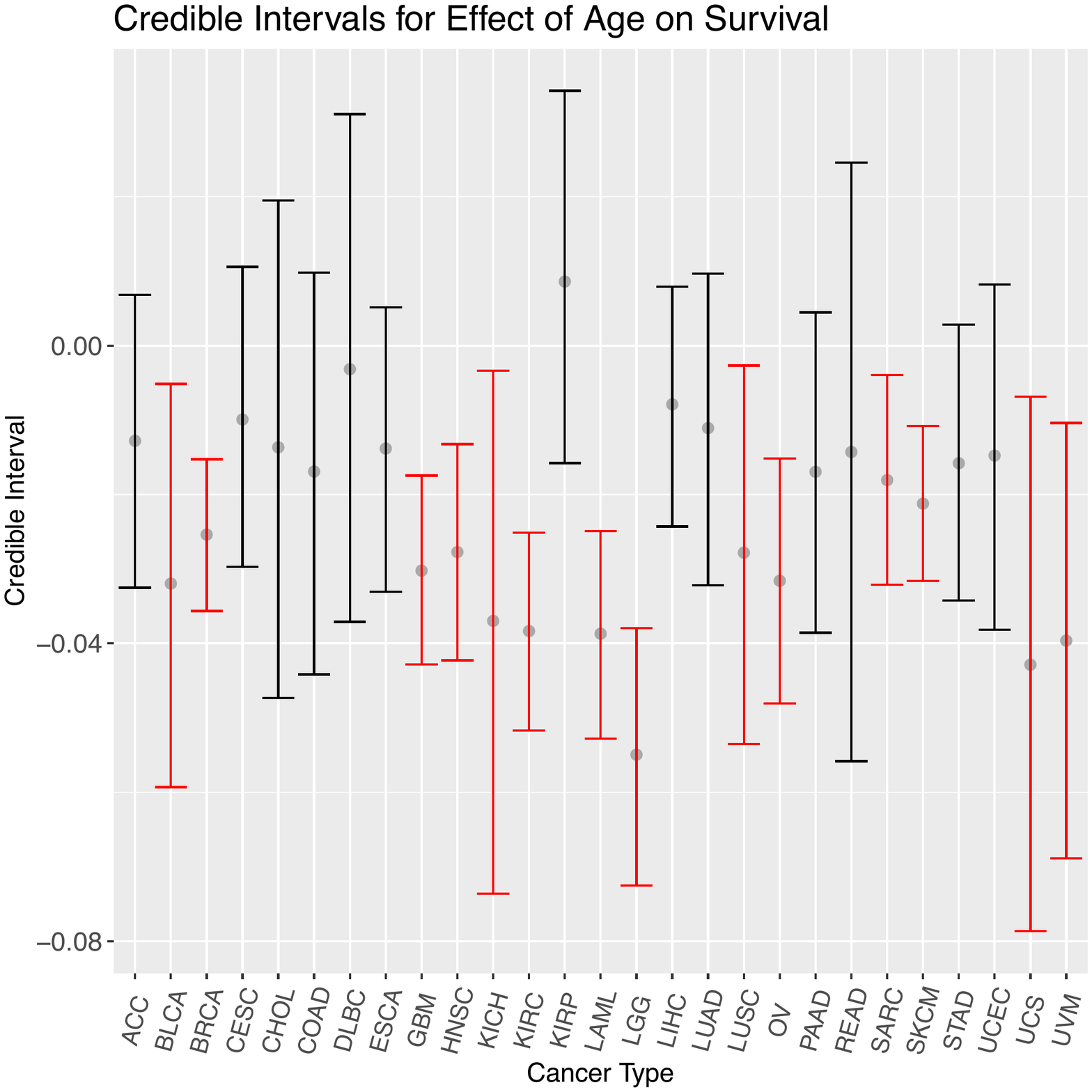}\label{fig:b}}

\vspace*{-1cm}
\centering
\sidesubfloat[]{\includegraphics[width=0.45\textwidth]{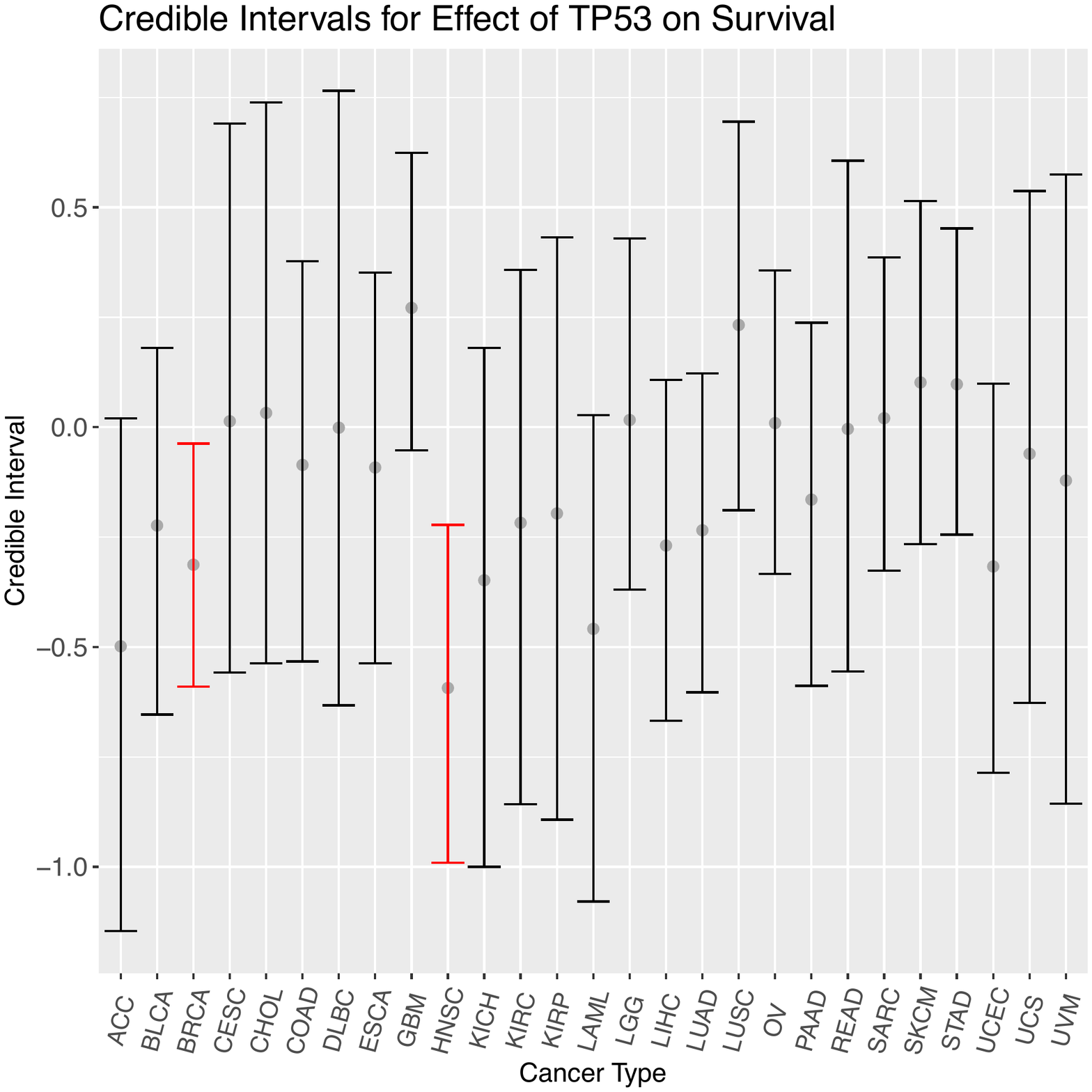}\label{fig:c}}
\hfil
\sidesubfloat[]{\includegraphics[width=0.45\textwidth]{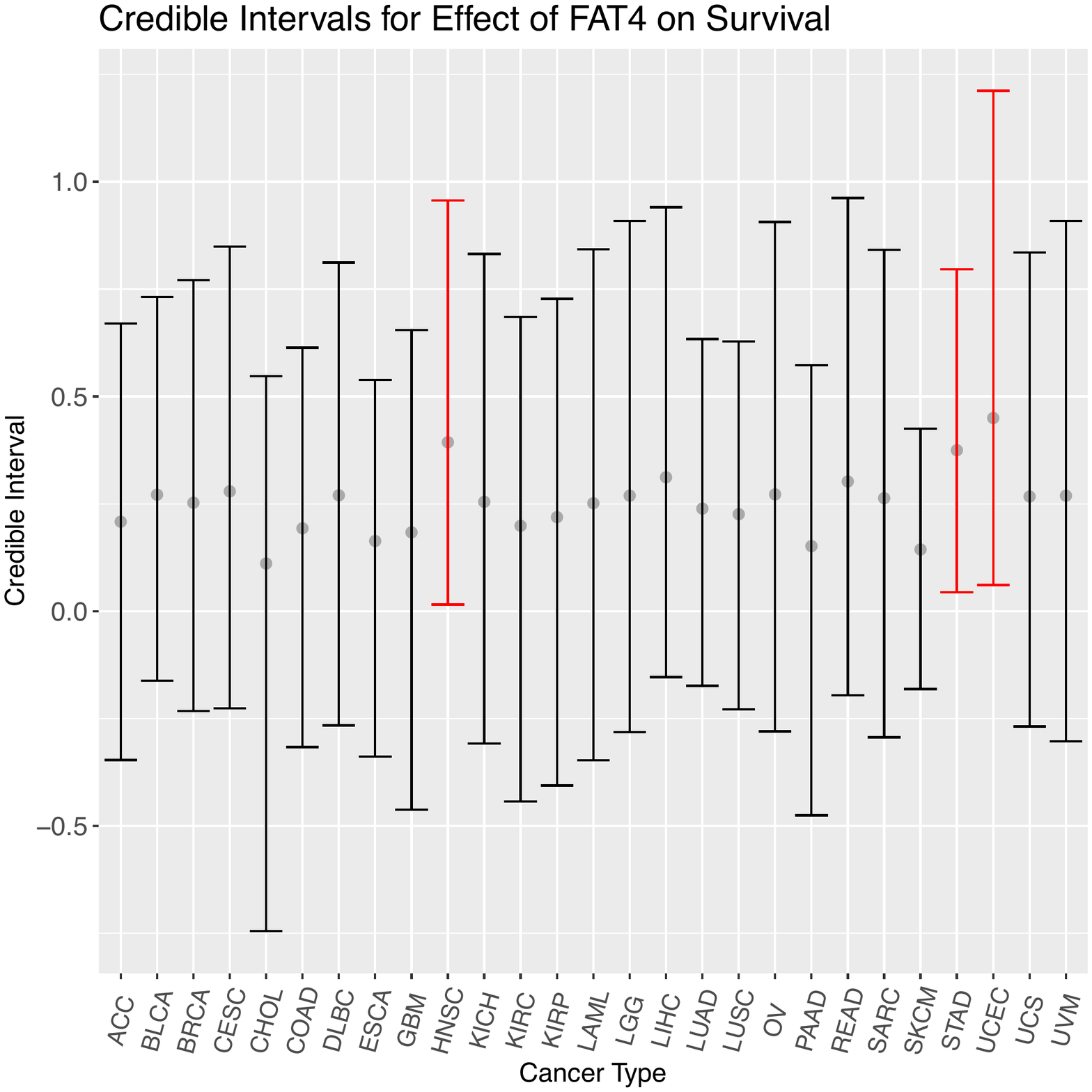}\label{fig:d}}

\caption{Credible Intervals By Covariate and Cancer Type}
\label{fig:myfigure}
\end{figure}

Figure~\ref{fig:myfigure} displays the credible intervals across cancer types for each parameter in the model. Panel~\ref{fig:a} compares the baseline survival across the different cancer types.  Panel~\ref{fig:b} reveals the generally deleterious effect of age on patient survival, as indicated by the highlighted red intervals. For the majority of cancers, an increase in age led to a decrease in survival; however, the extent to which age has an effect is not homogeneous or precisely identified for every cancer. For breast cancer (BRCA), the impact of age is more certain, indicated by a narrower credible interval, compared to the effect of age on patients with (e.g.) uterine carcinosarcoma (UCS). Similarly, Figure~\ref{fig:c} shows the estimated effect of a \emph{TP53} mutation on survival across cancers, and the estimated effect was generally negative for most cancers. This, again, demonstrates a poorer prognosis for patients with a mutation at \emph{TP53}. Breast cancer (BRCA) and head-neck squamous cell carcinoma (HNSC) had estimated \emph{TP53} effects with credible bounds entirely below 0, and adrenocortical carcinoma (ACC) was nearly entirely below 0. \emph{FAT4} mutation credible intervals (Figure~\ref{fig:d}) appeared to be more positive than those of \emph{TP53}, with some intervals entirely above 0. 

\begin{table}[H]
\centering
\caption{Credible Intervals for Model Coefficients Across Each Cancer Type}
\label{tab:cred_intervals}
\scalebox{0.85}{
\begin{tabular}{rlllll}
 & Cancer & Intercept & Age & \emph{FAT4} & \emph{TP53} \\ 
  \hline
  1 & ACC & (7.616, 8.347) & (-0.033, 0.007) & (-0.347, 0.67) & (-1.146, 0.020) \\ 
  2 & BLCA & (6.677, 7.379) & (-0.059, -0.005) & (-0.162, 0.732) & (-0.654, 0.180) \\ 
  3 & BRCA & (8.675, 9.021) & (-0.036, -0.015) & (-0.233, 0.771) & (-0.590, -0.037) \\ 
  4 & CESC & (7.757, 8.313) & (-0.030, 0.011) & (-0.226, 0.849) & (-0.558, 0.691) \\ 
  5 & CHOL & (6.402, 7.381) & (-0.047, 0.019) & (-0.745, 0.547) & (-0.537, 0.739) \\ 
  6 & COAD & (7.348, 8.134) & (-0.044, 0.010) & (-0.317, 0.614) & (-0.533, 0.377) \\ 
  7 & DLBC & (7.65, 8.721) & (-0.037, 0.031) & (-0.266, 0.812) & (-0.633, 0.765) \\ 
  8 & ESCA & (6.412, 7.284) & (-0.033, 0.005) & (-0.339, 0.539) & (-0.537, 0.351) \\ 
  9 & GBM & (5.657, 6.036) & (-0.043, -0.017) & (-0.462, 0.655) & (-0.053, 0.624) \\ 
  10 & HNSC & (7.225, 7.906) & (-0.042, -0.013) & (0.016, 0.957) & (-0.991, -0.222) \\ 
  11 & KICH & (8.662, 9.781) & (-0.074, -0.003) & (-0.308, 0.832) & (-1.000, 0.180) \\ 
  12 & KIRC & (7.724, 8.054) & (-0.052, -0.025) & (-0.443, 0.685) & (-0.858, 0.358) \\ 
  13 & KIRP & (7.941, 8.611) & (-0.016, 0.034) & (-0.406, 0.727) & (-0.893, 0.431) \\ 
  14 & LAML & (6.213, 6.666) & (-0.053, -0.025) & (-0.347, 0.843) & (-1.079, 0.028) \\ 
  15 & LGG & (7.676, 8.306) & (-0.072, -0.038) & (-0.282, 0.909) & (-0.369, 0.429) \\ 
  16 & LIHC & (6.761, 7.257) & (-0.024, 0.008) & (-0.154, 0.94) & (-0.668, 0.108) \\ 
  17 & LUAD & (7.188, 7.742) & (-0.032, 0.010) & (-0.174, 0.634) & (-0.603, 0.122) \\ 
  18 & LUSC & (6.549, 7.36) & (-0.054, -0.003) & (-0.229, 0.628) & (-0.189, 0.695) \\ 
  19 & OV & (6.963, 7.568) & (-0.048, -0.015) & (-0.280, 0.906) & (-0.334, 0.356) \\ 
  20 & PAAD & (6.302, 7.071) & (-0.039, 0.004) & (-0.476, 0.572) & (-0.588, 0.238) \\ 
  21 & READ & (7.376, 8.58) & (-0.056, 0.025) & (-0.196, 0.962) & (-0.556, 0.606) \\ 
  22 & SARC & (7.37, 7.844) & (-0.032, -0.004) & (-0.294, 0.842) & (-0.326, 0.386) \\ 
  23 & SKCM & (7.537, 7.927) & (-0.032, -0.011) & (-0.181, 0.425) & (-0.266, 0.515) \\ 
  24 & STAD & (6.495, 7.033) & (-0.034, 0.003) & (0.044, 0.796) & (-0.244, 0.452) \\ 
  25 & UCEC & (8.467, 9.097) & (-0.038, 0.008) & (0.061, 1.212) & (-0.786, 0.099) \\ 
  26 & UCS & (6.270, 7.518) & (-0.079, -0.007) & (-0.269, 0.835) & (-0.627, 0.537) \\ 
  27 & UVM & (7.268, 8.046) & (-0.069, -0.010) & (-0.303, 0.909) & (-0.856, 0.574) \\ 
   \hline
\end{tabular}
}
\end{table}

\subsection{Credible intervals for Mean Mutation Effect} 
\begin{figure}[H]
  \includegraphics[width=0.5\linewidth]{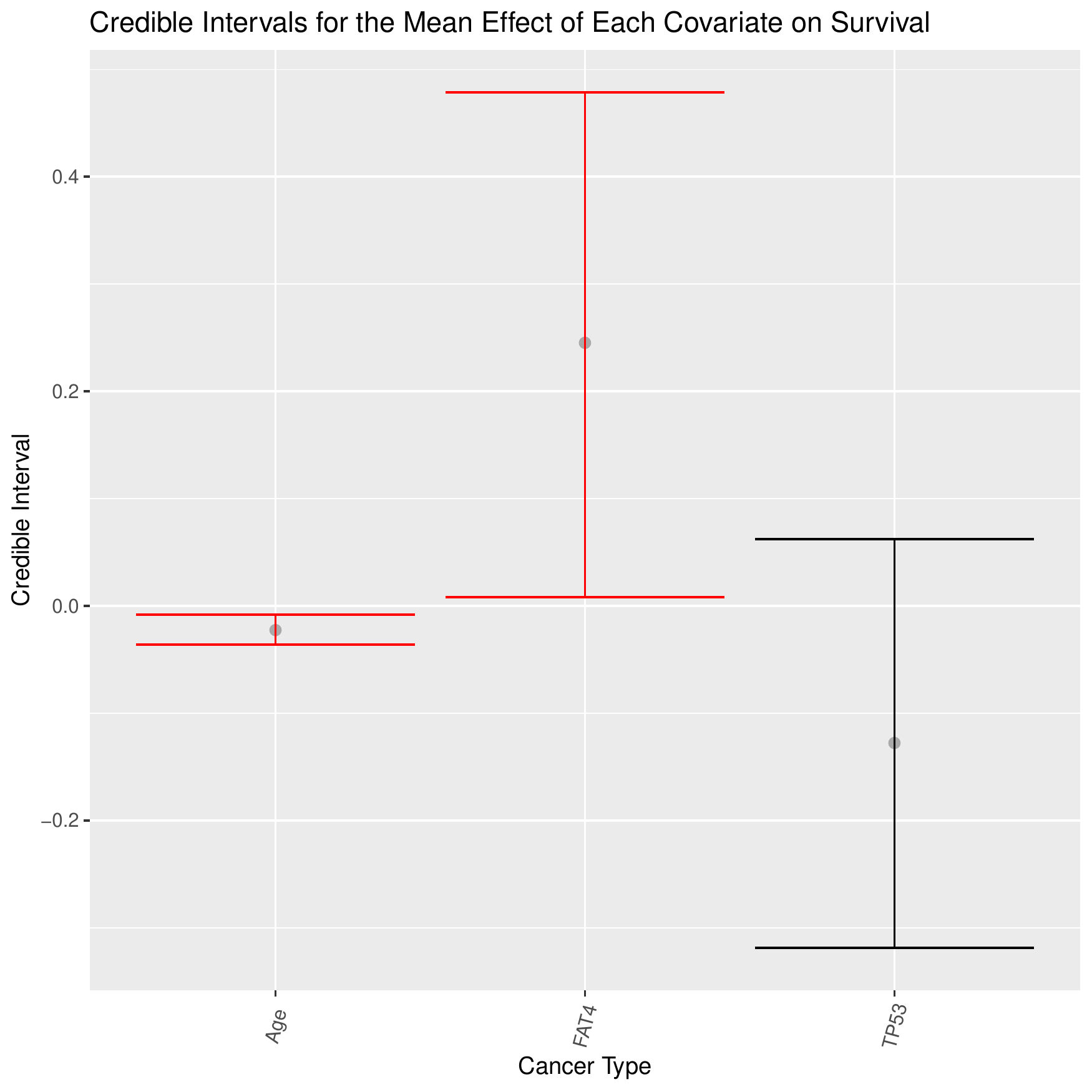}
  \caption{Visual display of the credible intervals for $\tilde\beta_p, \hspace{1mm} p = 1, \dots, 5$}
  \label{fig:mean_intervals}
  \label{fig: Beta Tilde Plot}
\end{figure}

We also studied the credible intervals for the mean effect of each covariate. The mean effect for each predictor, $\tilde\beta_p$, was assumed constant across cancer types and the individual effect by tumor varied around this mean. The results for $\tilde\beta_p$ are given in Table~\ref{tab:mean_intervals} and shown in Figure~\ref{fig:mean_intervals}. \\

\begin{table}[]
\centering
\caption{Credible Intervals for Mean of Model Coefficients}
\label{tab:mean_intervals}
\begin{tabular}{cc}
Covariate             & Mean Effect        \\ \hline
Intercept             & (7.213, 7.870)     \\
Age                   & (-0.037, -0.008)   \\
\emph{FAT4} Mutation Status  & (0.031, 0.480)    \\
\emph{TP53} Mutation Status & (-0.316, 0.056)    \\
\end{tabular}
\end{table}

These results indicate the effect of each covariate averaged across cancers; therefore, if a predictor was more potent in one cancer and less so in another, this may not necessarily be represented in estimates for the mean effect. This substantiates why we chose to allow the effect of each covariate to differentiate by tumor type. The credible intervals for age and \emph{TP53} mutation status coincide with the $\beta_{ip}$ interval results as both had entirely negative or nearly entirely negative interval estimates for their respective means. The interval for \emph{FAT4} also coincides with the $\beta_{ip}$ intervals; however, its comparably large width demonstrates a lack of certainty on its effect with age and \emph{TP53} in the model. 

\subsection{Survival Plots}
\begin{figure}[H]
  \includegraphics[width=1\linewidth]{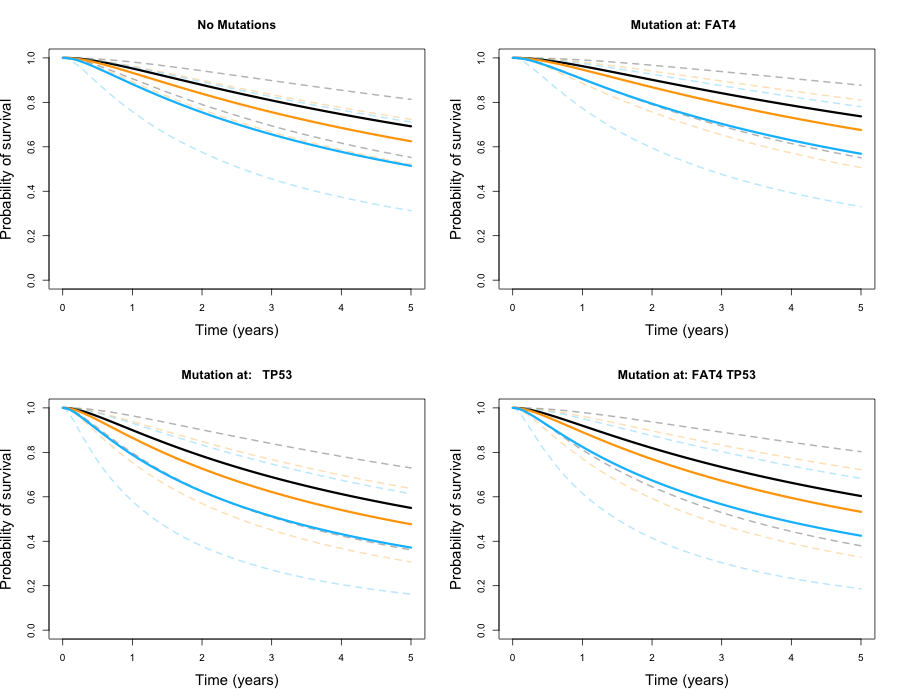}
  \caption{Survival curves under different covariate combinations for ACC with ages overlaid. Estimates for 30 year old patients are shown in black, 50 year olds in orange, and 80 year olds in blue. 95\% error bounds are shown in dotted lines.}
  \label{fig:acc_plots}
\end{figure}

To visualize the impact of age and a mutation at each of \emph{TP53} and \emph{FAT4}, we show here 
survival curves computed based on each combination of predictor values. The full collection of survival curves, for any cancer type and any combination of predictors, are available online at \url{http://ericfrazerlock.com/surv_figs/SurvivalDisplay.html}. The plots displayed in Figure~\ref{fig:acc_plots} are for a combination of covariates for patients with adrenocortical carcinoma (ACC), for which we had data on 89 tumors. In our dataset, patients with ACC ranged in age from 14 years old to 72 years old, with a median age of 49. The mutation rates for patients with ACC are as follows: 50.6\% had no mutations in \emph{FAT4} or \emph{TP53}, 10.1\% had mutations in just \emph{FAT4}, 19.1\% had mutations at just \emph{TP53}, and 4.49\% had mutations in both. The impact of the negative coefficient of \emph{TP53} is demonstrated in these plots, as prognosis seems to worsen over 5 years if a patient has such a mutation. The deleterious effect of age is also visible, which is to be expected. The seemingly positive effect of \emph{FAT4} is also apparent, with calculated survival curves appearing higher compared to those for patients with no mutations at \emph{TP53} and \emph{FAT4}. 

\section{Discussion}
\label{discussion}
In this article, we propose a novel Bayesian hierarchical model for cancer patient survival based on age and mutation status. This model is unique in its ability to allow the effect of each covariate to vary by cancer type. This framework is motivated by the assumption that similar genetic profiles may have similar, though not necessarily identical, effects on patient survival across tissue-of-origin. This work may be extended to allow for other clinical covariates to be added to the model, such as stage and grade, and allows the user to adjust the effect of each predictor by cancer type as informed by prior knowledge.

To determine which genes were most important in survival prediction, we used a forward selection procedure that added \emph{TP53} and \emph{FAT4} to our model.  The inclusion of \emph{TP53} led to a dramatic improvement in the model fit, while each additional gene reduced the log-posterior likelihood by a marginal amount. This indicates that \emph{TP53} is largely the most predictive, which is natural given its high mutation rate across cancer types (see Table~\ref{tab:threeparts}). In particular, robust effects for \emph{TP53} were observed for breast cancer (BRCA) and head-neck squamous cell carcinoma (HNSC); the basal-like subtype for BRCA\cite{cancer2012comprehensive} and the HPV-negative subtype for HNSC\cite{cancer2015comprehensive} are almost universally \emph{TP53}-mutated and have relatively poor outcomes.  Moreover, there is a vast literature on the mechanistic role of \emph{TP53} in cancer progression as an agent of DNA repair \cite{olivier2010TP53} and in maintenance of genome integrity.   It was also encouraging to see \emph{TP53} added to the model as it is considered a tumor driver gene\cite{bailey2018comprehensive}.  \emph{FAT4} appears to be much less predictive than \emph{TP53}, given its marginal increase in the log-posterior likelihood (from -1009.63 to -1009.135). It is of note that the credible intervals for coefficient in the model were largely positive, despite the existence of literature concluding \emph{FAT4} functions as a tumor suppressor \cite{Cai2015FAT4, Wei2019FAT4}. A potential explanation is that mutations in \emph{FAT4} contribute to the development of certain cancers, but these cancers are comparatively less aggressive than those that arise from mutations in \emph{TP53} or other driver genes.      An analysis without \emph{TP53} achieved comparable predictive performance under our hierarchical model via the inclusion of \emph{\emph{APOB}}, suggesting that even genes with lower mutation rates can improve performance. Using the most frequently mutated genes across genome sequencing cohorts often also includes known false positives.  Comparing our list of genes with Bailey et al\cite{bailey2018comprehensive}, we found that 19/50 were on known false positive lists. However, it was encouraging to see that these known false positives did not make the final survival model. However, our gene set also did not include some known tumor driver genes due to our approach of restricting the genes of interest to those with mutations across the PanCancer cohort.   

In addition, we considered the possibility of collinearity between mutation status variables, prompting us to investigate the correlation levels between variables across all cancer types (Figure~\ref{fig:Correlation_Plot}). Based on this plot, we concluded mutation statuses across all cancer types were not highly associated. However, the results may look different if one is not considering the cancers in aggregate. With that in mind, we propose in future work sorting genes differently to meet the interest of the researchers, i.e. exploring genes that are known to be highly mutated in a set of related cancers. In such an instance, it may be necessary to consider collinearity between variables and adjust accordingly. 

We also assessed the predictive quality of our model by calculating survival curves for each cancer type based on combinations of age and mutation status based on our model. These plots demonstrated the largely negative effect of increasing age and a \emph{TP53} mutation and the less noticeable effect of \emph{FAT4}. Irrespective of mutation status, the survival curves demonstrate the clear effect age has on survival prognosis, which does not come as much of a surprise. We created a web-based app  that allows users to toggle with cancer types and predictors to see what 5-year prognosis is predicted to be. Such a tool may be interesting for academic purposes only. 

In our study, we used the TCGA2STAT R package to import TCGA somatic mutation data due to its convenience in data dissemination, providing somatic mutation data in a ready-to-use format for statistical analysis. While the package does offer convenience, we were not able to acquire data for Mesothelioma, which is available through the TCGA Multi-Center Mutation-Calling in Multiple Cancers (MC3) dataset and through the NCI's Genome Data Commons. In the interest of ease of future replication of this study, using the TCGA2STAT dataset may make it simpler for scientists to acquire the same data we used. We also did not distinguish the genes we used based on driver mutation status, false positive status, or otherwise.

In future studies, it would be interesting to apply the model to a subset of the 27 cancer types we selected in order to group cancers that may be more similar in genetic nature or otherwise. This may elucidate genes unique to predicting patient survival outcome to specific cancer groupings. Interactions between covariates could also be included in the model to assess their relationship to overall survival. It would also be interesting to investigate alternate approaches to selecting genes to incorporate in the model, possibly incorporating prior knowledge on driver mutation status or false positive status. 

\section{Acknowledgements}
This work was supported by the National Institutes of Health (NIH) National Cancer Institute (NCI) grant R21CA231214-01.

\appendix
\section{Appendix 1: Additional Model Fitting Algorithm Details} 
We used an in-house Gibbs sampler to estimate the parameters of our proposed log-normal and normal survival models. At each iteration of our sampler, we drew a sample of a parameter from its respective conditional posterior distribution. The conditional posterior distributions for each parameter are outlined below: 

\begin{gather*}
    \vec{\beta_{i}} \mid (X_i, y_i, \tilde\beta, \sigma^2, \lambda^2) \sim \mbox{Normal} ([\frac{1}{\sigma^2} X_i^T X_i + \frac{1}{\lambda^2} I]^{-1}[\frac{1}{\sigma^2}X_i^T y_i + \frac{1}{\lambda^2}\tilde\beta], \\ \frac{1}{\sigma^2} X_i^T y_i +  \frac{1}{\lambda^2}\tilde\beta)
\end{gather*}

where $\vec{\beta_i}$ is the vector of coefficients for the $i$th cancer type, $X_i$ is the design matrix for the $i$th cancer type, $y_i$ is the vector of survival times for the $i$th cancer type with the censored observations replaced by their time-of-last-contact, $\tilde\beta$ is the vector of coefficient parameter means, and $\lambda^2$ is the vector of variances for each coefficient parameter.\cite{Lindley1972Linear}

\[
    \lambda_p \mid (\vec{\beta_i}, \tilde\beta) \sim \mbox{Inverse-Gamma}(\frac{K}{2} + 0.01, 0.01 + \frac{1}{2}W) 
\]

for $p = 1, \dots, P$, $P$ being the number of coefficients in the model, $i = 1, \dots, K$ and where $K = 27$, the number of cancer types and $W = \sum_{i=1}^{27} (\vec{\beta_i} - \tilde\beta)^2$. 

\[
    \tilde\beta_p \mid (\vec{\beta_i}, \lambda_p^2) \sim \mbox{Normal}(\frac{K \tau^2 \Bar{\beta}_p}{\lambda_p^2 + K\tau^2}, \frac{\lambda_p^2\tau^2}{\lambda_p^2 + K\tau^2})
\]

where $\Bar{\beta}_p$ is the average coefficient for covariate $p$ across all cancer types, $\tau^2 = 10000^2$. 

\[
    \sigma^2 \mid (X_i, y_i) \sim \mbox{Inverse-Gamma}(\frac{N}{2} + 0.01, \frac{1}{2}B + 0.01)
\]

where $N$ is the total number of observations in the model and $B = \sum_{i=1}^K (y_i - X_i \vec{\beta_i})^2$. 

\section{Appendix 2: Validation Study}
In fitting the normal and log-normal models, we ran our own Gibbs sampling algorithm to generate posteriors for each of the parameters. To validate that our sampler was running properly, we generated true values for each parameter which we used to generate simulated data. Using this data, we computed posteriors for each parameter, calculated 95\% credible intervals, and checked whether the true value of the parameter was contained within the interval. The entire algorithm in more detail is outlined below: 

\renewcommand{\theenumii}{\arabic{enumii}}
\begin{enumerate}
    \item For $i$ in $1, \dots, 1000$ iterations, 
    \begin{itemize}
        \item Initialize a counter at 0 to store the number of iterations out of 1000 for which a parameter has been contained in its calculated credible interval.  
        \item Generate values of predictors from $N(0,1)$ to be stored in matrix $X$ and generate "true" values for each parameter, $\vec{\beta_0}, \vec{\tilde\beta_0}, \lambda^2_0, \sigma^2_0$ from its respective prior distribution.
            \begin{itemize}
                \item Based on initial values of parameters, generate survival times from a normal distribution with mean $X\vec{\beta_0}$ and variance $\sigma^2_0$. Generate censor times from the same distribution.
                \item Replace survival times for observations whose censor time is less than its survival time with NAs to indicate that that observation has been censored. 
            \end{itemize}
        \item Using simulated data, run the algorithm as described in section 2. We chose to generate 2000 posterior samples. 
        \item Once samples have been generated, calculate 95\% credible intervals for each parameter.
        \begin{itemize}
            \item Using the "true" values, check if the calculated credible interval covers the true value of the parameter. If so, update the counter for this parameter by one. 
        \end{itemize}
    \end{itemize}
    \item Ensure that for approximately 95\% of the 1000 iterations, the credible interval covered the true value of the parameter. 
\end{enumerate}

\bibliographystyle{SageV}
\bibliography{bibliography}

\end{document}